\documentclass[11pt]{article}
\pdfoutput=1 

\usepackage{jheppub} 
\usepackage{hyperref}

\subheader{UTTG-16-16}

\title{Noether charge, black hole volume, and complexity}

\author[a]{Josiah Couch,}
\author[a]{Willy Fischler,}
\author[a]{and Phuc H. Nguyen}

\affiliation[a]{Theory Group, Department of Physics and Texas Cosmology Center, University of Texas\\
Austin, TX 78712, USA}

\emailAdd{josiah.couch@utexas.edu}
\emailAdd{fischler@physics.utexas.edu}
\emailAdd{phn229@physics.utexas.edu}

\abstract{In this paper, we study the physical significance of the thermodynamic volumes of AdS black holes using the Noether charge formalism of Iyer and Wald. After applying this formalism to study the extended thermodynamics of a few examples, we discuss how the extended thermodynamics interacts with the recent complexity = action proposal of Brown et al. (CA-duality). We, in particular, discover that their proposal for the late time rate of change of complexity has a nice decomposition in terms of thermodynamic quantities reminiscent of the Smarr relation. This decomposition strongly suggests a geometric, and via CA-duality holographic, interpretation for the thermodynamic volume of an AdS black hole. We go on to discuss the role of thermodynamics in complexity = action for a number of black hole solutions, and then point out the possibility of an alternate proposal, which we dub ``complexity = volume 2.0". In this alternate proposal, the complexity would be thought of as the spacetime volume of the Wheeler-DeWitt patch. Finally, we provide evidence that, in certain cases, our proposal for complexity is consistent with the Lloyd bound whereas CA-duality is not.}

\begin{document}

\maketitle
\flushbottom

\section{Introduction}
The laws of black hole thermodynamics, at least in their traditional formulation \cite{Bekenstein:1973ur, Bekenstein:1974ax, Hawking:1974sw, Hawking:1976de}, do not include a pressure-volume conjugate pair. This conspicuous absence is perhaps related to the difficulty of defining the volume of a black hole in a coordinate-invariant way: unlike the area of the horizon, a na\"ive integration over the interior of a black hole depends on the foliation of spacetime. A number of relativists \cite{Kastor:2009wy, Dolan:2012jh, Kubiznak:2014zwa, Frassino:2015oca, Kubiznak:2016qmn}, and more recently high energy physicists \cite{Johnson:2014yja}, have suggested that the pressure should be identified as the cosmological constant. In this framework, dubbed the extended black hole thermodynamics or ``black hole chemistry'', the ADM mass of the black hole is reinterpreted as the enthalpy $H$ of the system rather than its internal energy $U$. The volume, then, can be defined in the usual thermodynamical way to be:
\begin{equation}\label{Vthermo}
    V = \left( \frac{\partial H}{\partial P} \right)_{S}.
\end{equation}
Fascinatingly enough, in simple cases such as the AdS-Schwarzschild or AdS-Reissner-Nordstrom (AdS-RN) black hole, the thermodynamic volume coincides with a na\"ive integration over the ``black hole interior'':

\begin{equation}\label{SchwarzschildVolume}
    V = \frac{4}{3}\pi r_{+}^{3} = \int_{0}^{r_{+}} \sqrt{-g} dr d\Omega_{2}^{2}.
\end{equation}

In more complicated cases, such as rotating holes or solutions with hair, the thermodynamical volume is less intuitive, and it is an interesting question to ask how the volume arises as an integral of some local quantity over some region of spacetime, in a way similar to (\ref{SchwarzschildVolume}). For a selection of work on or related to this topic, we refer to \cite{Caceres:2015vsa, Karch:2015rpa, Caceres:2016xjz, Nguyen:2015wfa, Kastor:2016bph, Kastor:2014dra,Pradhan:2016rff}.

Our main goal in this paper is two-fold: on the one hand, we attempt to shed light on the meaning of the thermodynamic volume as a geometrical quantity, which, a priori, is an abstract notion of volume associated to the black hole and does not correspond to the actual volume of any spatial region. On the other hand, we will relate the thermodynamic volume to holography, and in particular to the quantum complexity of the boundary state. Why should we believe that the thermodynamic volume has a role to play in the holographic context? To this question, we will offer four answers, which we list one after another below.

The first reason to believe that the thermodynamic volume has a place in holography is that, as we will demonstrate below, this quantity is derivable from the Noether charge (or Iyer-Wald) formalism \cite{Iyer:1994ys, Iyer:1995kg}, or a slight twist thereof. This is the main finding of section \ref{Sec:IyerWald}. A powerful way to derive the first law of black hole thermodynamics, the Iyer-Wald formalism has yielded deep insights into the nature of black hole entropy, so it is a natural step to extend the formalism to derive the thermodynamic volume. In recent years, the Iyer-Wald formalism has proved useful to holographers as a means to translate between the geometry in the bulk to quantum information theoretic quantities on the boundary, starting with \cite{Faulkner:2013ica} where the formalism was used to derive the linearized equation of motion in the bulk from the first law of entanglement on the boundary in pure AdS. To give a few more examples, the formalism was used in \cite{Lin:2014hva} to relate matter in the bulk to the relative entropy on the boundary, in \cite{Lashkari:2015hha} to relate canonical energy in the bulk to the quantum Fisher information on the boundary, in \cite{Lashkari:2016idm} to relate quantum information inequalities to gravitational positive energy theorems, and finally in \cite{Mosk:2016elb} in conjunction with the kinematic space program to clarify the emergence of gravity from entanglement.

The second reason to suspect that the thermodynamic volume is relevant to holography is that various notions of volume in the bulk have been identified with quantum information theoretic quantities. In particular, the size of the Einstein-Rosen bridge of a 2-sided eternal AdS black hole is believed to capture the complexity of the thermofield double state \cite{Brown:2015bva, Brown:2015lvg}. Furthermore, the complexity of subregions of the boundary CFT \cite{Alishahiha:2015rta,Momeni:2016ira,Momeni:2016ekm, Ben-Ami:2016qex} and the fidelity \cite{Miyaji:2015mia} have been related to the volume of a constant time slice in the bulk enclosed between the Ryu-Takayanagi surface and the boundary. In the light of these ideas, it is suggestive that the thermodynamical volume also admits a quantum information theoretic interpretation, and we will find in this paper that it indeed seems so. In the second part of the paper (sections \ref{Sec:ComplexityI} and \ref{Sec:ComplexityII}), we will relate the thermodynamic volume (and also the Smarr relation) to the complexity as per the proposals in \cite{Brown:2015bva, Brown:2015lvg}.

The third motivation to study the thermodynamic volume in holography is the question of to what extent holography knows about the black hole interior. In \cite{Hartman:2013qma}, the black hole interior was probed with minimal surfaces which cross the horizon, and the notion of ``vertical entanglement'' as well as a tensor network picture of black hole interiors were formulated. Like the quantum complexity, the vertical entanglement could serve as a useful way to keep track of time in the dual CFT. In the case of the complexity, the key observation is that the size of the ERB grows linearly with the boundary time at late time, a behavior conjectured to be true of the quantum complexity at exponentially late time.

However this linear growth can be observed for various geometrical entities crossing the wormhole, and to correctly pick out one among them is a non-trivial problem. This leads us to the fourth and last motivation of this paper: how to correctly quantify the size of an ERB and capture the complexity? Two ways to achieve this have been considered in the literature \cite{Susskind:2014rva, Stanford:2014jda, Brown:2015bva,Brown:2015lvg}. The first way, first proposed in \cite{Stanford:2014jda} and dubbed ``complexity=volume'' or CV-duality, postulates that the complexity is dual to the volume of the maximal spatial slice crossing the ERB. This proposal, while capturing the linear growth at late time, has the minor problem that a length scale has to be introduced ``by hand'', for which there seems to be no unique, natural choice. The second way, first proposed in \cite{Brown:2015bva} and dubbed ``complexity=action'' or CA-duality, postulates that the complexity is dual to the bulk action evaluated on the Wheeler-DeWitt (WDW) patch. This proposal solves the length scale problem of CV-duality and has, in addition, the practical advantage that the WDW patch is easier to work with than the maximal volume. We will see in this paper that the thermodynamic volume is intimately related to the linear growth of the WDW patch at late time. We will also point out a third possibility, dubbed ``Complexity = Volume 2.0" in which complexity is identified with the spacetime volume of the WDW patch rather than the action. This is potentially even easier to work with and will be discussed in more detail further in the paper. 

The paper is organized as follows: In Section \ref{Sec:IyerWald}, we briefly review the Iyer-Wald formalism (with varying cosmological constant) and apply it to derive the thermodynamic volume of two solutions: the charged BTZ black hole and the R-charged black hole. In Section \ref{Sec:ComplexityI}, we move on to discuss the connection between the extended thermodynamics and the complexity in the simple case of the AdS-Schwarzschild black hole. In Section \ref{Sec:ComplexityII}, we extend this connection with the complexity to a black hole with conserved charges (i.e. electrically charged and rotating holes). In Section \ref{ActionvsVolume}, we contrast our proposal for the complexity against CA-duality and show that, in certain cases, our proposal can help fix problems which ail CA-duality. In Section \ref{Sec:Conclusion}, we conclude and discuss future work.

\section{Volume and Iyer-Wald Formalism}\label{Sec:IyerWald}
In this section we will present a slight generalization of the Iyer-Wald formalism \cite{Iyer:1994ys,Iyer:1995kg,Jacobson:1993vj} which will allow us to derive the volume. The formalism requires a diffeomorphism invariant action $S=\int \mathbf{L} = \int \mathcal{L} \boldsymbol{\epsilon}$, together with a solution with a bifurcate timelike Killing vector field $\xi$. We will follow the notation used in \cite{Faulkner:2013ica} \footnote{In particular, $\boldsymbol{\epsilon}$ denotes the usual volume form in $d$ dimensions:
\begin{equation}
\boldsymbol{\epsilon} = \frac{1}{d!} \sqrt{-g} \text{ } \varepsilon_{a_1..a_d} dx^{a_1} \wedge ... \wedge dx^{a_d}
\end{equation}
We will find it useful to define the additional forms:
\begin{gather*}
\boldsymbol{\epsilon}_\mu = \frac{1}{(d-1)!} \sqrt{-g} \text{ } \varepsilon_{\mu a_1 ... a_{d-1}} dx^{a1} \wedge ... \wedge dx^{a_{d-1}}\\
\boldsymbol{\epsilon}_{\mu \nu} = \frac{1}{(d-2)!} \sqrt{-g} \text{ } \varepsilon_{\mu \nu a_1 ... a_{d-2}} dx^{a1} \wedge ... \wedge dx^{a_{d-2}}
\end{gather*}
}. Consider some general variation of the Lagrangian. For an action including a cosmological constant which is allowed to vary, one writes:
\begin{equation}
\delta \mathbf{L} = d\boldsymbol{\Theta} + \sum_{\phi} \mathbf{E}^{\phi} \delta \phi + \frac{\partial \mathbf{L}}{\partial \Lambda} \delta \Lambda,
\end{equation}
where
\begin{equation}
\boldsymbol{\Theta}(\delta \phi) = \sum_\phi \frac{\partial \mathbf{L}}{\partial \left( \nabla_\mu \phi \right)} \delta \phi
\end{equation}
is the symplectic potential current,
\begin{equation}
\mathbf{E}^{\phi} = \frac{\partial \mathbf{L}}{\partial \phi} - \nabla_\mu \frac{\partial \mathbf{L}}{\partial \left( \nabla_\mu \phi \right)}
\end{equation}
is the equation of motion form for the field $\phi$, and where the sum over $\phi$ runs over the entire field content of the theory. In the case where this variation is due to applying a diffeomorphism generated by a vector field $\zeta$, this becomes
\begin{equation}
    \delta_\zeta \mathbf{L} = d\boldsymbol{\Theta}(\delta_\zeta \phi) + \sum_\phi \mathbf{E}^\phi \delta_\zeta \phi.
\end{equation}
In this case, since our action is diffeomorphism invariant, we may apply Noether's theorem to derive the conserved current 
\begin{equation}
    \mathbf{J}(\zeta) = \boldsymbol{\Theta}(\delta_\zeta \phi) - \zeta\cdot \mathbf{L}
\end{equation}
and a Noether charge form $\mathbf{Q}(\zeta)$ such that on-shell
\begin{equation}
    \mathbf{J}(\zeta) = d\mathbf{Q}(\zeta).
\end{equation}
Replacing our general vector field $\zeta$ by the killing vector field $\xi$, and considering some general other variation $\delta \phi$, we now define 
\begin{equation}
    \boldsymbol{\chi} = \delta_\phi \mathbf{Q}(\xi) - \xi\cdot \boldsymbol{\Theta}(\delta \phi).
\end{equation}
By an algebraic computation we may see that on-shell
\begin{equation}
    d\boldsymbol{\chi} = -\xi \cdot \frac{\partial \mathbf{L}}{\partial \Lambda} \delta \Lambda.
\end{equation}
Applying Stokes' theorem to this form on a region $\Sigma$ of a constant time slice bounded by the bifurcate killing horizon and the conformal boundary at infinity then yields
\begin{equation}\label{eq:ExtendedIyerWald}
    \int_\Sigma d\boldsymbol{\chi} = - \delta \Lambda \int_\Sigma \xi \cdot \frac{\partial \mathbf{L}}{\partial \Lambda} = \int_\infty \boldsymbol{\chi} - \int_\mathcal{H} \boldsymbol{\chi}.
\end{equation}
In the case of a black hole spacetime, this reduces to the extended first law of black hole thermodynamics upon evaluation of the integrals, where roughly speaking, the second term from the left above gives rise to the $VdP$.\\

\subsection{Application to Einstein-Maxwell: charged BTZ black hole}
Next, we apply the above formalism to the Einstein-Maxwell theory:
\begin{equation}
S = \int d^d x \sqrt{-g} \left[\left(R - 2\Lambda \right) - \frac{1}{4}F^2 \right].
\end{equation}
After some algebra, we find the symplectic potential current, Noether current and Noether charge to be:
\begin{gather*}
\boldsymbol{\Theta} = \Theta^\mu \boldsymbol{\epsilon}_\mu \\
\mathbf{J} = J^\mu \boldsymbol{\epsilon}_\mu \\ 
\mathbf{Q} = Q^{\mu\nu} \boldsymbol{\epsilon}_{\mu\nu}
\end{gather*}
with
\begin{gather*}
\Theta^\mu = \left[ 2\left(\nabla_\nu \nabla^{[\nu}\xi^{\mu]} + g^{\mu\nu} R_{\nu\lambda}\xi^\lambda \right) - F^{\mu\nu}\left( \nabla_\nu (\xi^\lambda A_\lambda) + \xi^\lambda F_{\lambda \nu} \right) \right] \\
J^\mu = \nabla_\nu \left[ 2 \nabla^{[\nu}\xi^{\mu]} -  F^{\mu\nu}\xi^\lambda A_\lambda \right] \\
Q^{\mu\nu} =  \left[\nabla^{[\nu}\xi^{\mu]} - \frac{1}{2} F^{\mu\nu}\xi^\lambda A_\lambda \right].
\end{gather*}
Let us now specialize to a solution of the Einstein-Maxwell system: the charged BTZ black hole in 3 dimensions. The metric together with the gauge field are given by:
\begin{gather*}
ds^2 = -f(r) dt^2 + \frac{dr^2}{f(r)} + r^2 d\phi^2\\
f(r) = -2m + \frac{r^2}{L^2} - \frac{1}{2} q^2 \log \left( \frac{r}{L} \right)\\
A = -q \log \left( \frac{r}{L} \right) dt.
\end{gather*}
Here we use units where $4 G = 1$. This solution has two horizons, an outer horizon at $r = r_+$ and an inner horizon at $r= r_-$. Both outside the outer horizon ($r > r_+$) and inside the inner horizon ($r < r_-$), $\partial_t$ is a bifurcate time-like Killing vector field, whose killing horizons are given by $r=r_+$ and $r=r_-$ respectively. For this choice of Killing vector field, we find that the Noether charge only has one nonzero component:
\begin{equation}
Q^{tr} = \frac{4 r^2 - 2 L^2 q^2 \log \left(\frac{r}{L}\right) - L^2 q^2}{16 \pi  L^2 r} 
\end{equation}
and all other independent components of $Q^{\mu\nu}$ vanish. For a perturbation defined by perturbing the AdS length $L$ of the solution above and leaving the other parameters fixed, we find that
\begin{equation}
\Theta^r = \frac{3 \left(4 r^2 - L^2 q^2\right)}{8 \pi  L^3 r} \delta L 
\end{equation}
with the other two components zero. From these we find the only nonzero component of $\chi$ to be:
\begin{equation}
\chi^{tr} = \frac{4 r^2 - L^2 q^2}{16 \pi  L^3 r} \delta L.
\end{equation}
Integrating this form over any surface of constant $t$ and $r$ yields
\begin{equation}\label{intOfChi}
\int_{r=\text{const}} \boldsymbol{\chi} = \int_0^{2\pi} \sqrt{-g} \chi^{\mu\nu} \varepsilon_{\mu\nu \phi} d\phi = \frac{\delta L}{L} \left[ \frac{r^2}{L^2} - \frac{1}{4} q^2 \right].
\end{equation}
Evaluating this on the outer horizon $r_+$ we can recongnize this as $T\delta S$ after some algebra. On the other hand, the integral of $\boldsymbol{\chi}$ diverges as $r\rightarrow \infty$, but putting in a large $r$ cutoff $R$ and adding
\begin{equation}
\delta \Lambda \int \xi \cdot \frac{\delta \mathbf{L}}{\delta \Lambda} = \frac{\delta L}{4 \pi}\int_0^{2\pi} d\phi \int_{r_+}^R dr \sqrt{-g} \frac{-4}{L^3} = \frac{r_+^2 - R^2}{L^3} \delta L 
\end{equation}
we get a cutoff independent result also equal to $T\delta S$. Rewriting this in terms of $\delta P$ instead of $\delta L$, the first law with $m$ and $q$ fixed then becomes
\begin{equation}
T \delta S + \pi \left(r_+^2 - \frac{1}{4} L^2 q^2 \right) \delta P = 0
\end{equation}
and so we have the volume
\begin{equation}\label{OuterBTZVol}
V_+ = \pi \left(r_+^2 - \frac{1}{4} L^2 q^2 \right)
\end{equation}
We could equally well have done this in region inside the inner horizon. Evaluating equation \ref{intOfChi} at $r=r_-$ once again yields $T\delta S$ for this horizon, and evaluating at the singularity gives $-\frac{q^2}{4L} \delta L$. On the other hand,
\begin{equation}
\delta \Lambda \int \xi \cdot \frac{\delta \mathbf{L}}{\delta \Lambda} =  \frac{\delta L}{4 \pi}\int_0^{2\pi} d\phi \int_{r_-}^0 dr \sqrt{-g} \frac{-4}{L^3} = \frac{r_-^2}{L^3} \delta L 
\end{equation}
All in all, we obtain the first law:
\begin{equation}
\left( T\delta S \right)_- + \frac{\delta L}{L} \left( \frac{r_-^2}{L^2} - \frac{1}{4} q^2 \right) = 0
\end{equation}
where the $(\dots)_{-}$ is to emphasize that the quantity enclosed pertains to the inner horizon. Once again trading $\delta L$ for $\delta P$ we read off the volume for the inner horizon:
\begin{equation}\label{InnerBTZVol}
V_- = \pi \left( r_-^2 - \frac{1}{4} q^2 L^2 \right)
\end{equation}

\subsection{Application to Einstein-Maxwell-dilaton: the R-charged Black Hole}
In this subsection, we consider a more complicated example and derive the volume of the R-charged black hole in 4 dimensions. The thermodynamics of R-charged black holes has been studied in \cite{Cvetic:2010jb}. In (3+1) dimension, the action is given by:
\begin{equation}
    L =  \left( \frac{R}{16\pi} - \frac{1}{8\pi} \sum_{I=1}^{4} e^{\vec{a}_{I} \cdot \vec{\phi}} F_{(I)}^{2} - \frac{1}{32\pi} \sum_{i=1}^{3} ((\partial \phi_{i})^{2}   - \mathcal{V}{(\phi_{i})} )\right) \varepsilon
\end{equation}
with
$$a_1 = (1,1,1) \text{, } a_2 = (1,-1,-1) \text{, } a_3 = (-1,1,-1) \text{, } a_4 = (-1,-1,1)$$ and
\begin{equation}
    \mathcal{V}{(\phi_{i})} = -\frac{g^{2}}{4\pi} \sum_{i} \cosh{\phi_{i}}
\end{equation}
The metric together with the matter fields are given by:
\begin{equation}
    ds^{2} = -\prod_{I=1}^{4}H_{I}^{-1/2} f dt^{2} + \prod_{I=1}^{4}H_{I}^{1/2} \left(\frac{dr^{2}}{f} + r^{2}d\Omega^{2} \right)
\end{equation}
\begin{equation}\label{VectorFieldRcharged}
    A^{I} = \frac{\sqrt{q_{I}(q_{I}+2m)}}{r+q_{I}} dt
\end{equation}
\begin{equation}\label{ScalarFieldRcharged}
    e^{-\frac{1}{2} \vec{a}_{I} \cdot \vec{\phi}} =  \frac{\prod_{J=1}^{4} H_{J}^{1/4}}{H_{I}}
\end{equation}
\begin{equation}
    f = 1 - \frac{2m}{r} + g^{2}r^{2} \prod_{J} H_{J}
\end{equation}
\begin{equation}
    H_{J} = 1 + \frac{q_{J}}{r}
\end{equation}
The thermodynamical quantities are:
\begin{equation}\label{ADMMassSTU}
    M = m + \frac{1}{4} \sum_{I=1}^{4} q_{I}
\end{equation}
\begin{equation}\label{ChargesSTU}
    Q_{I} = \frac{1}{2} \sqrt{q_{I}(q_{I}+2m)}
\end{equation}
\begin{equation}
    S = \pi \prod_{I=1}^{4} \sqrt{r_{+}+q_{I}}
\end{equation}
\begin{equation}
    T = \frac{f'(r_{+})}{4\pi} \prod_{I=1}^{4} H_{I}^{-1/2}{(r_{+})}
\end{equation}
\begin{equation}
    \Phi^{I} = \frac{\sqrt{q_{I}(q_{I}+2m)}}{2(r_{+}+q_{I})}
\end{equation}
In the extended phase space, the pressure is the cosmological constant, which is also the bottom of the scalar potential:
\begin{equation}
    P = \frac{3}{8\pi} g^{2}
\end{equation}
As mentioned in the introduction, the ADM mass is now reinterpreted as the enthalpy and the black hole's volume can be computed using the familiar thermodynamic formula:
\begin{equation}\label{RchargedVolume}
V := \left( \frac{\partial M}{\partial P}\right)_{Q_I,S} = \frac{\pi}{3} r_{+}^{3} \prod_{J=1}^{4} H_{J}{(r_{+})} \sum_{K=1}^{4} H_{K}{(r_{+})}^{-1}   
\end{equation}
In particular, the AdS-RN black hole is a special case when all 4 charges coincide $q_{1}=q_{2}=q_{3} \equiv q$. In this case, the above reduces to:
\begin{equation}
    V = \frac{4}{3}\pi (q+r_{+})^{3}
\end{equation}
Also, the radial coordinate has to be redefined by $r \rightarrow r - q$ in order to recover the usual Schwarzschild-like form of the AdS-RN metric. We then recognize the volume of the AdS-RN black hole in the form of equation (\ref{SchwarzschildVolume}). A note here is in order about coordinate dependence. While the thermodynamic volume can take different forms depending on the radial coordinate used (as illustrated in the example above), we stress that the volume is coordinate-invariant quantity. The fact that it is not the actual volume of some spatial region, combined with the fact that spatial volumes in General Relativity depend on the foliation, can make this coordinate invariance not so obvious. The cleanest way to see this coordinate invariance is to go back to the definition of $V$ as a partial derivative (\ref{Vthermo}). The function $M{(S,P)}$, which represents an equation of state so to say, is a relation between coordinate invariant quantities ($M$, $S$ and $P$), and so is the partial derivative (\ref{Vthermo}).\\
The paper \cite{Cvetic:2010jb} asks the interesting question of what integral over the black hole interior would give rise to the volume (\ref{RchargedVolume}). To answer this question, one can recast the above in the form:
\begin{equation}\label{VolumeToScalarPotential}
V = \int_{S^{2}} \int_{r_{0}}^{r_{+}} V'{(r)} dr d\Omega_{2}^{2}
\end{equation}
where $V(r)$ is the function defined in equation (\ref{RchargedVolume}) (with $r_{+}$ relabeled to $r$), and $r_{0}$ is taken to be the largest root of the equation $V(r)=0$. We then find that $r_{0}$ is the largest root of a cubic polynomial:
\begin{equation}
4r_{0}^{3} + 3r_{0}^{2} \sum_{I} q_{I} + 2 r_{0} \sum_{i<j} q_{I} q_{J} + \sum_{I<J<K} q_{I} q_{J} q_{K} = 0
\end{equation}
As for the integral $V'(r)$, it was pointed out in \cite{Cvetic:2010jb} that it is essentially the scalar potential:
\begin{equation}\label{VolumeRchargedIntegral}
    V =  -\frac{8\pi}{3g^{2}} \int_{S^{2}} \int_{r_{0}}^{r_{+}} \mathcal{V} \sqrt{-g} dr d\Omega_{2}^{2}
\end{equation}
Two aspects of this formula are remarkable: first, the fact that the integrand admits a clean interpretation in terms of the scalar potential; and secondly, the integral does not run over the whole of the black hole's interior. As one can generally expect the volume to have something to do with the scalar potential, the second aspect is perhaps a bit more mysterious than the first one. We now proceed to apply the extended Iyer-Wald formalism to compute the volume, and we will see how the formalism sheds light on the two mysterious aspects as described above. The symplectic potential current and Noether charge for this theory are given by:
\begin{equation}
\Theta^{a} = \nabla_{b} (g^{ad}g^{bc}\delta g_{cd} - g^{ab}g^{cd} \delta g_{cd}) - \sum_{i=1}^{3} \nabla^{a}\phi_{i} \delta \phi_{i} - 8 \sum_{I} e^{\vec{a}_{I} \cdot \vec{\phi}} F_{(I)}^{ab} (\partial_{b}(\xi^{c}A_{(I)c}) + \xi^{c} F_{(I)cb} )
\end{equation}
\begin{equation}
Q^{ab} = -\frac{1}{16\pi}\nabla^{[b}\xi^{a]} + \frac{1}{4\pi} \sum_{I} e^{\vec{a}_{I} \cdot \vec{\phi}} F_{(I)}^{ab} \xi^{c} A_{(I)c}
\end{equation}
Here we only give the on-shell form of these expressions. Next, let us perturb the coupling $g^{2}$. By noting that equations (\ref{VectorFieldRcharged}) and (\ref{ScalarFieldRcharged}) are $g$-independent, it is clear that the profiles of the matter fields are unaffected, and only the gravity part contributes to $\delta Q$ and $\Theta$. Moreover, equations (\ref{ADMMassSTU}) and (\ref{ChargesSTU}) are also $g$-independent, so neither the ADM mass $M$ nor the charges $Q_{i}$ are affected by the $g$-variation. This implies that the (extended) first law of thermodynamics:
\begin{equation}
\delta M = T \delta S + \Phi_{I} dQ_{I} + VdP
\end{equation}
simplifies to
\begin{equation}
T\delta S + V \delta P = 0
\end{equation}
If we now compare with equation (\ref{eq:ExtendedIyerWald}), we can identify the $T\delta S$ term with the integral of $\chi$ over the horizon, and the $V \delta P$ term as arising from a combination of the two other terms. The fact that $T\delta S$ corresponds to the integral of $\chi$ over the horizon is to be expected from the Iyer-Wald formalism: roughly speaking, it is because the form $\chi$ evaluated on the bifurcation surface reduces to the surface binormal, and hence its integral over the bifurcation surface gives the area (or the entropy).\\
Let us next compute the form $\chi$. After some algebra, we find that the only nonzero component of $\chi$ is:
\begin{equation}
\chi^{rt} = -\frac{1}{16\pi} \left(\frac{r}{2}\right) \left( 2\sqrt{H_{1}H_{2}H_{3}H_{4}} + r\partial_{r}{\sqrt{H_{1}H_{2}H_{3}H_{4}}} \right) \delta g^{2}
\end{equation}
The integral of $\chi$ over infinity diverges. If we regularize by a radial cutoff $r_{c} >> r_{+}$, we find:
\begin{equation}\label{int1}
\int_{\infty} \chi = \left[ -\frac{r_{c}^{3}}{2} - \frac{3r_{c}^{2}}{8} \sum_{I} q_{I} - \frac{r_{c}}{4} \sum_{I<J} q_{I}q_{J} - \frac{1}{8} \sum_{I<J<K} q_{I}q_{J}q_{K} \right] \delta g^{2}
\end{equation}
Next, let us focus on the $\delta \Lambda$ term in the extended first law. By differentiating the Lagrangian with respect to the coupling $g^{2}$, we have:
\begin{equation}
\delta g^{2} \int \frac{\partial \mathcal{L}}{\partial g^{2}} \xi \cdot \varepsilon = -\frac{\delta g^{2}}{g^{2}} \int_{S^{2}} \int_{r_{+}}^{\infty} \mathcal{V} \sqrt{-g} dr d\Omega_{2}^{2}
\end{equation}
Notice that we have an integral of the scalar potential on the right-hand side! We emphasize here that the extended Iyer-Wald formalism makes this fact manifest, in contrast with the approach described in equation (\ref{VolumeToScalarPotential}). As usual, the upper limit of integration above will diverge and we have to regularize by a radial cutoff $r_{c}$. Evaluating the integral, we then find:
\begin{equation}\label{int2}
\delta g^{2} \int \frac{\partial \mathcal{L}}{\partial g^{2}} \xi \cdot \varepsilon = \delta g^{2} \left[ \frac{r^{3}}{2} + \frac{3}{8}r^{2} \sum_{I} q_{I} + \frac{1}{4}r \sum_{I<J} q_{I}q_{J} \right]_{r_{+}}^{r_{c}}
\end{equation}
If we now compare the divergent terms in (\ref{int1}) and (\ref{int2}), we then find that they cancel pairwise, and we are left with a finite answer which consists of two parts: (1) the finite term in (\ref{int1}) and the horizon term (the lower limit of integration) in (\ref{int2}). We then obtain:
\begin{eqnarray}
V dP &=& \int_{\infty} \chi + \delta g^{2} \int \frac{\partial \mathcal{L}}{\partial g^{2}} \xi \cdot \epsilon \nonumber \\
&=& \delta g^{2} \left(\frac{r_{+}^{3}}{2} + \frac{3}{8}r_{+}^{2} \sum_{I}q_{I} + \frac{r_{+}}{4} \sum_{I<J} q_{I}q_{J} - \frac{1}{8} \sum_{I<J<K}q_{I}q_{J}q_{K} \right)
\end{eqnarray}
and we recover equation (\ref{RchargedVolume}). Notice in particular, that, from the viewpoint of the extended Iyer-Wald formalism, the lower limit of integration $r_{0}$ in (\ref{VolumeRchargedIntegral}) arises from the finite term in the integral of $\chi$ at infinity. Moreover, the Iyer-Wald formalism has taught us that the volume of the black hole is perhaps best thought of as arising from an integral over the {\it exterior} of the black hole rather than its {\it interior} \footnote{We also note here that there exists an alternative approach in the literature to derive the volume, which is based on the Komar formula, see e.g. \cite{Frassino:2015oca, Cvetic:2010jb, Kastor:2009wy}. As with our approach, the volume also arises from the Komar formula as a integral over the exterior of the black hole, and the integrand is basically the Killing potential.}. To summarize, the volume arises as the integral of the scalar potential over the whole black hole exterior, but it is regularized by the Iyer-Wald form $\chi$ at infinity in a nontrivial way.\\

\section{Thermodynamic volume and Complexity: Schwarzschild-AdS}\label{Sec:ComplexityI}
From the viewpoint of the Iyer-Wald formalism, as we have seen above, the black hole volume arises as an integral over the {\it exterior} of the black hole. This observation naturally begs the question of whether the thermodynamic volume has something to do with the black hole {\it interior}. Moreover, it remains unclear as to what the information contained in the volume can teach us about the dual CFT. It is generally pointed out in the literature \cite{Kastor:2009wy, Caceres:2016xjz, Kastor:2016bph, Kastor:2014dra, Johnson:2014yja, Caceres:2015vsa, Dolan:2013dga, Kubiznak:2016qmn} that varying the cosmological constant in the bulk corresponds to varying the rank of $SU(N)$ or the central charge on the field theory side, and that the volume can be thought of as a chemical potential-like quantity corresponding to the degrees of freedom counted by the central charge.\\

In this section, we bring the two questions above together (the black hole interior and the CFT interpretation) and attempt to answer them through the notion of {\it complexity} of quantum states. In a series of elegant papers \cite{Brown:2015bva, Brown:2015lvg}, Brown et al. proposed that the complexity of the CFT state is dual to the integral of the bulk action over the so-called Wheeler-DeWitt (WdW) patch. In particular, this quantity grows linearly with time at late time, and we will see in the first half of this section that the thermodynamical volume is a contribution to this growth. In the section half of this section, we switch gear and consider the possibility that it is the \textit{spacetime volume}, rather than the action, of the WDW patch which is dual to the complexity. We will show that the spacetime volume of the WDW patch is intimately related to the thermodynamic volume, and that, in the Schwarzschild-AdS case, the spacetime volume and action behave in very similar fashions and both proposals should work equally well.\\

\subsection{Review of Brown et al.}
\begin{figure*}[h!]
$$
 \includegraphics[width=8cm]{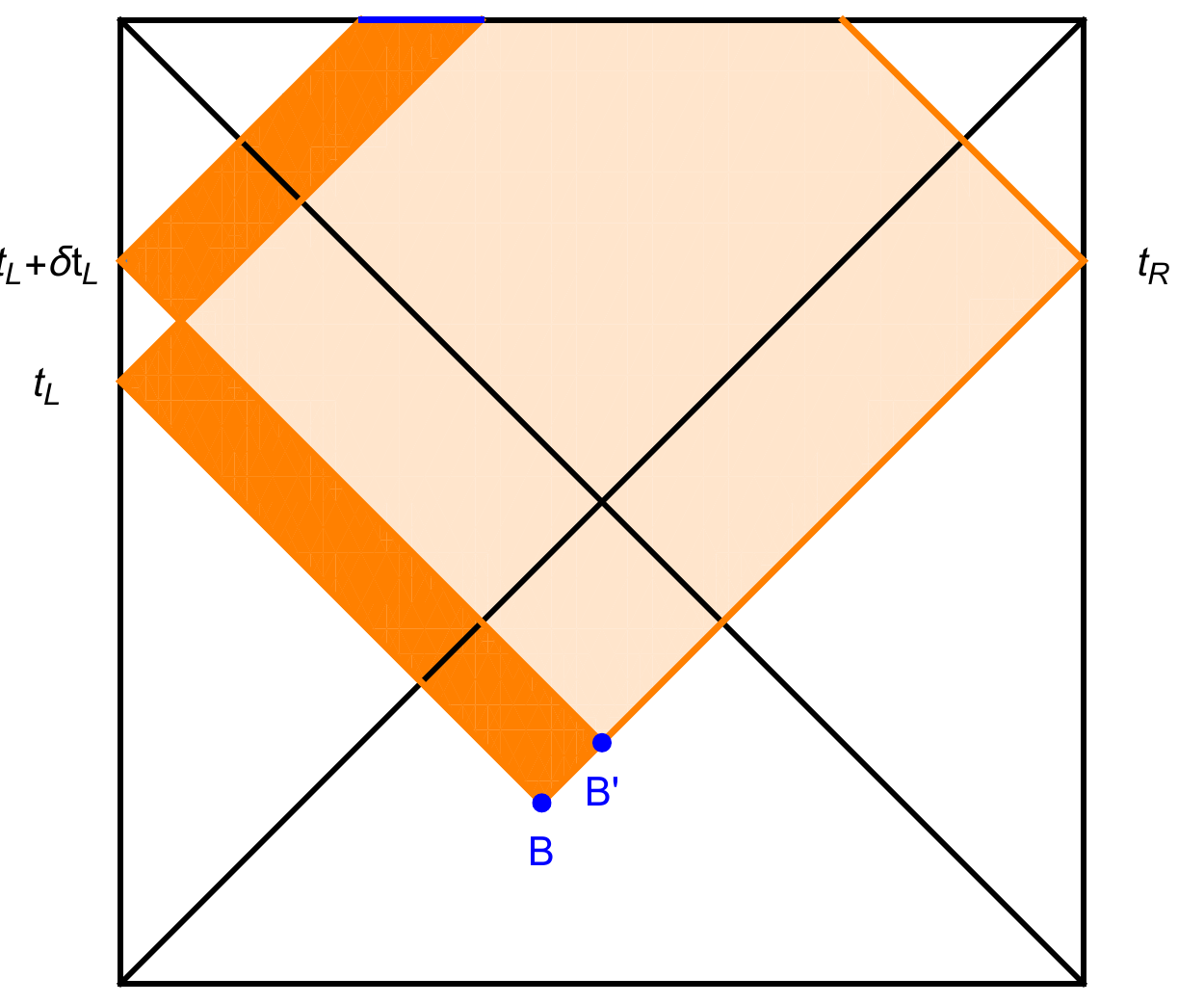}
$$
\caption{The Wheeler-DeWitt patch of the AdS-Schwarzschild black hole (depicted in orange). When $t_{L}$ is shifted to $t_{L} + \delta t_{L}$, the patch loses a sliver and gains another one (depicted in darker orange). The contributions from the Gibbons-Hawking term are in blue.}
\label{Fig:WdWpatchAdSSchw}
\end{figure*}

Let us start by reviewing the proposal by Brown et al. in some level of details. The Wheeler-DeWitt patch is a region in the maximally extended black hole spacetime defined with respect to two choices of time, one on each boundary. For simplicity let us first consider the AdS-Schwarzschild black hole in 4 dimensions. We will denote the time on the left boundary as $t_{L}$ and the time on the right boundary as $t_{R}$. From these two points on the boundary (see Figure \ref{Fig:WdWpatchAdSSchw} for a depiction), we draw four null rays, and the WdW patch is the region in the bulk enclosed between rays (and possibly the past and future singularities) \footnote{The WdW patch as described here extends all the way to the boundary, and therefore the action evaluated on the WdW is divergent. To extract a finite answer, we have to choose a regularization. One could simply cut off the patch at some large radius $r_{\mathrm{cutoff}} >> r_{+}$. Alternatively, one could move the two corners of the WdW patch on the boundary to $r_{\mathrm{cutoff}}$, as done in \cite{Lehner:2016vdi}. The regularization introduces terms which drop out when we take the time derivative of the complexity, and for this reason, we leave the regularization unspecified.}.\\
On the CFT side, picking out two times $t_{L}$ and $t_{R}$ is equivalent to choosing a quantum state:
\begin{equation}
| \psi{(t_{L},t_{R})} \rangle = e^{-i(H_{L}t_{L} + H_{R}t_{R})} |TFD \rangle
\end{equation}
where $H_{L}$ and $H_{R}$ are the Hamiltonian on the left and right boundaries, respectively, and $| TFD \rangle$ is the thermofield double state:
\begin{equation}
| TFD \rangle = Z^{-1/2} \sum_{n} e^{-\beta E_{n}/2} |E_{n} \rangle_{L} \otimes |E_{n} \rangle_{R}
\end{equation}
The thermofield double state has the properties that it is close to being maximally entangled, and that the reduced density matrix on either side is the usual thermal state.

The complexity of a quantum state is, roughly speaking, the minimal number of quantum gates needed to produce the state from some universally agreed-upon starting point. The statement of CA-duality is that:
\begin{equation}
\mathcal{C}{(| \psi{(t_{L},t_{R})} \rangle)} = \frac{\mathcal{A}}{\pi \hbar}
\end{equation}
where $\mathcal{A}$ is the bulk action evaluated on the WDW patch. At late $t_{L}$, it follows from CA-duality that the rate of growth of the complexity approaches the mass of the black hole:
\begin{equation}\label{latetimegrowth}
\lim_{t_{L} \rightarrow \infty} \frac{d\mathcal{C}}{d t_{L}} = \frac{2M}{\pi \hbar}
\end{equation}
Equation (\ref{latetimegrowth}) is a convincing piece of evidence for CA-duality. This is because it is reminiscent of a conjectured upper bound on the rate of computation by Lloyd (\cite{Lloyd}), according to which the rate of computation is bounded above by the energy. Let us briefly review the motivation for the Lloyd bound.

The Lloyd bound takes inspiration from another bound known as the Margolus-Levitin theorem \cite{Margolus:1997ih}. This latter states that the time $\tau_{\perp}$ it takes for a quantum state to evolve into a state orthogonal to it is bounded below by:
\begin{equation}
    \tau_{\perp} \geq \frac{h}{4 E}
\end{equation}
where $E$ is the average energy of the state. If we take the reciprocal of both sides and re-interpret the left-hand side (which has unit of frequency) as the rate of change of the complexity, we then arrive at the statement that this rate is bounded above by the energy of the system:
\begin{equation}
\dot{\mathcal{C}} \leq  \frac{2E}{\pi \hbar}
\end{equation}
which is the Lloyd bound. We point out that, while the Margolus-Levitin theorem can be proved using elementary techniques, the Lloyd bound is a {\it conjecture}. If we now compare the Lloyd bound with the prediction of CA-duality (
\ref{latetimegrowth}) for the late-time complexification rate, we see that the ADM mass of the black hole plays the role of energy in the Lloyd bound and that the bound is {\it saturated} at late time. That the bound is saturated is another {\it conjecture} but is appealing: black holes seem to excel at information-related tasks since they saturate the Bekenstein bound \cite{Bekenstein:1980jp, Bekenstein:2004sh} and are believed to be the fastest scramblers in nature \cite{Sekino:2008he}.

\subsection{CA-duality, through the lens of black hole chemistry}
In this subsection, we take a closer look at the gravity calculation of CA-duality to derive equation (\ref{latetimegrowth}). This computation itself, of course, can be found in \cite{Brown:2015bva,Brown:2015lvg} \footnote{A technical remark is in order here. The method of computation in \cite{Brown:2015bva, Brown:2015lvg} was questioned by \cite{Lehner:2016vdi}, where the calculation was redone with a more careful treatment of the boundary of the WDW patch. However the conclusion \ref{latetimegrowth} remains unchanged. In this section, we will follow the more rigorous treatment of the boundary term as presented in \cite{Lehner:2016vdi}.}. Our contribution in this subsection is to show that the thermodynamic volume (together with the pressure) arises naturally from the calculation.

First, since the WdW patch is a region with boundary, the action is the sum of the Einstein-Hilbert action and the Gibbons-Hawking(-York) term:
\begin{equation}
    \mathcal{A} = \frac{1}{16\pi G} \int_{\mathcal{M}} \sqrt{-g} (R - 2\Lambda) + \frac{1}{8\pi G} \int_{\partial M} \sqrt{|h|} K
\end{equation}
When we shift $t_{L}$ to $t_{L} + \delta t_{L}$, the WdW patch loses a thin rectangle and gains another thin rectangle as described in dark orange in Figure \ref{Fig:WdWpatchAdSSchw}. Thus, to compute the rate of change of the action we have to evaluate the action above on the two orange rectangles. Observe that all the sides of these two rectangles are null except at the singularity, and the paper \cite{Lehner:2016vdi} gives a detailed argument that the null boundaries do not contribute to the Gibbons-Hawking term. Also, since the boundary is not smooth at the corners of the rectangles, we have to take into account the contributions localized at these corners (named $B$ and $B$' in Figure \ref{Fig:WdWpatchAdSSchw}). Thus, we see that the Gibbons-Hawking term contributes at the singularity, at $B$ and at $B'$ (all of which are depicted in blue in Figure \ref{Fig:WdWpatchAdSSchw}):

\begin{equation}\label{ContributionstodeltaA}
\delta \mathcal{A} = S_{\mathcal{V}_{1}} - S_{\mathcal{V}_{2}} - \frac{1}{8\pi G} \int_{S} K d\Sigma + \frac{1}{8\pi G} \oint_{B'} a dS - \frac{1}{8\pi G} \oint_{B} a dS
\end{equation}

Note that $\mathcal{V}_1$ and $\mathcal{V}_2$ denote the upper and lower dark orange slivers from figure \ref{Fig:WdWpatchAdSSchw} respectively, and that $a = \ln |k\cdot \bar{k}|$ where $k$ and $\bar{k}$ are the null normals to the corner pieces. Let us consider first the difference between the two rectangles $S_{\mathcal{V}_{1}} - S_{\mathcal{V}_{1}}$. Note that the Ricci scalar of the AdS-Schwarzschild solution is a constant:
\begin{equation}
R = \frac{2d}{d-2}\Lambda
\end{equation}
This readily follows from the fact that AdS-Schwarzschild is a vacuum solution of Einstein-Hilbert theory. Thus, if we evaluate the Einstein-Hilbert action on the AdS-Schwarzschild background, we immediately see that we have something proportional to the {\it spacetime volume}:
\begin{equation}
S_{\mathcal{V}_{1}} - S_{\mathcal{V}_{2}} \propto \int_{\mathcal{V}_{1}} \sqrt{-g} d^{4}x - \int_{\mathcal{V}_{2}} \sqrt{-g} d^{4}x
\end{equation}
Thus, after one evaluates the integrals above, we expect to see something which is schematically the product of a spatial volume and the infinitesimal time interval $\delta t$:
\begin{equation}\label{SV1-SV2Schematically}
S_{\mathcal{V}_{1}} - S_{\mathcal{V}_{2}} = \mathrm{(some~spatial~volume)} \delta t
\end{equation}
Let now us do the integral for $S_{\mathcal{V}_{1}} - S_{\mathcal{V}_{2}}$ explicitly. When we do this, two remarkable things happen. The is that the part of the upper rectangle which is outside the future horizon always cancels with the part of the lower rectangle which is outside the past horizon, and this happens for any $t_{L}$ thanks to boost symmetry of the black hole \footnote{Indeed the Schwarzschild-AdS metric in Kruskal coordinates only depends on the coordinates $U$ and $V$ through their product $UV$, and is therefore invariant under $U \rightarrow e^{\beta} U$ and $V \rightarrow e^{-\beta}V$.}. Thus, whatever quantity comes out to be the spatial volume in equation (\ref{SV1-SV2Schematically}) only receives contribution from the black hole interior. The second is that the integral evaluates to:
\begin{equation}
S_{\mathcal{V}_{1}} - S_{\mathcal{V}_{2}} = -\frac{r_{B}^{3}}{2G L^{2}} \delta t
\end{equation}
where $r_{B}$ is the $r$ coordinate of the 2-sphere sitting at $B$. In the late time limit, we can easily see by inspection of Figure (\ref{Fig:WdWpatchAdSSchw}) that $r_{B}$ tends to $r_{+}$. Thus, in the late time limit, the integral above can be interpreted in the language of the extended thermodynamics as:
\begin{equation}\label{ComplexityAndVolume}
S_{\mathcal{V}_{1}} - S_{\mathcal{V}_{2}} = -\frac{r_{+}^{3}}{2GL^{2}} \delta t = -PV \delta t
\end{equation}
where, in the last equality, we used $P = -\frac{\Lambda}{8\pi G}$, $\Lambda = -\frac{3}{L^{2}}$ and $V = \frac{4}{3}\pi r_{+}^{3}$. Thus, we have seen how the thermodynamic volume arises from the action evaluated on the WDW patch. Put differently, the WDW patch provides an interpretation of hte thermodynamic volume as a measure of the black hole interior, and in the same time, relates it to the late-time rate of growth of the complexity.

Let us now evaluate the remaining contributions in (\ref{ContributionstodeltaA}) (The algebraic details are found again in \cite{Lehner:2016vdi}). The contribution of the Gibbons-Hawking term at the singularity is essentially the ADM mass:
\begin{equation}
-\frac{1}{8\pi G} \int_{S} K d\Sigma = \frac{3}{2}M \delta t
\end{equation}
As for the contribution at the two corners $B$ and $B'$, one finds:
\begin{equation}
\frac{1}{8\pi G} \left[ \oint_{B'} a dS - \oint_{B} a dS \right] = \frac{1}{4G} \left[ r^{2} \frac{df}{dr} + 2r f \log{\left(\frac{-f}{K}\right)} \right]_{r =r_{B}} \delta t
\end{equation}
where $K$ is a constant. In the late time limit, where $r_{B} \rightarrow r_{+}$, the second term above vanishes and:
\begin{equation}
\frac{1}{8\pi G} \left[ \oint_{B'} a dS - \oint_{B} a dS \right] = T S \delta t
\end{equation}
Putting everything together, we find the time derivative of the action at late time to be:
\begin{equation}
\frac{d\mathcal{A}}{dt} = \frac{3}{2}M + TS - PV
\end{equation}
Next, recall the Smarr relation for AdS-Schwarzschild in 4 dimensions:
\begin{equation}
M = 2TS - 2PV
\end{equation}
Using the Smarr relation above, the time derivative of the action simplifies to:
\begin{equation}
\frac{d\mathcal{A}}{dt} = 2M
\end{equation}
If we now turn the logics around, we can reinterpret the following slight rewriting of the Smarr relation:
\begin{equation}
2M = \frac{3}{2}M + TS - PV
\end{equation}
as a way to keep track of the different contributions to the complexity growth: the left-hand side corresponds to the total growth, the term with $M$ on the right-hand side is the contribution from the singularity of the WdW patch, the term with $TS$ is the corner contributions which end up on the horizon at late time, and finally the term with $PV$ is the contribution from the black hole interior away from the singularity.

\subsection{Complexity = Volume 2.0}\label{Sec:CVdualityv2}
As we have learned from (\ref{ComplexityAndVolume}), in the case of AdS-Schwarzschild, the late-time rate of change of the bulk action evaluated on the WDW patch gives the product $PV$, or equivalently the late-time rate of change of the spacetime volume of the WDW patch is the thermodynamic volume $V$. These observations beg the question of whether $P$ and $V$ can serve as the basis for a new, alternative proposal for the complexity alongside with CA-duality. In this subsection, we will make the case that a possible holographic dual to the complexity is the spacetime volume of the WDW patch.

As previously noted, what we are looking for in proposing a holographic dual to the complexity is a linear growth at late time, together with consistency with the Lloyd bound. On the information-theoretical side, the linear growth of the complexity at late time is generally believed to be true but is surprisingly hard to prove \footnote{``Late time'' in this statement refers to timescales exponential in the entropy. For much later times (doubly exponential in the entropy), quantum recurrence kicks in, and the complexity periodically returns to zero.}. It is straightforward to see that the complexity of the thermofield double state is bounded above by a linear function of time:
\begin{equation}\label{UpperBoundLinearGrowth}
\mathcal{C}{(|\psi{(t)}\rangle)} < t \cdot \mathrm{poly}{(K)}
\end{equation}
almost by definition of the complexity. To see this, recall that the complexity is the smallest number of quantum gates needed to build a state, hence any way to build the state automatically establishes an upper bound on the complexity. In particular, time-evolving the thermofield double state the usual way in quantum mechanics establishes the upper bound (\ref{UpperBoundLinearGrowth}). To establish that the complexity grows linearly at late time, one needs to also bound the complexity from {\it below} by a linear function of time. This is a highly non-trivial task, but there are two promising directions. One of them is a recently proved theorem by Aaronson and Susskind \cite{Aaronson:2016vto} which establishes a lower bound for the complexity (modulo the possibility that an improbable statement in complexity theory is true). The other direction is Nielsen's idea of the complexity geometry \cite{DowlingNielsen} where finding the complexity reduces to the problem of finding geodesics on a curved manifold.

On the gravity side, as noted in the introduction already, one can associate various geometrical quantities to the ERB which all grow linearly in size at late time, so this property of the complexity alone allows for quite some freedom in proposing a holographic dual. A simple illustration of this non-uniqueness phenomenon (given in \cite{Stanford:2014jda}) is a geodesics in the BTZ black hole anchored at boundary times $t_{L}$ and $t_{R}$. The length of such a geodesic is given by (for the case $r_{+}=L$):
\begin{equation}
    d{(t_{L},t_{R})} = 2\log{\left(\cosh{\frac{1}{2}(t_{L}+t_{R})}\right)}
\end{equation}
If we keep $t_{R}$ fixed and send $t_{L} \rightarrow \infty$, we find that indeed the leading term is linear in $t_{L}$. Another geometrical entity whose size grows linearly at late time is the maximal surface spanning the wormhole. As previously noted, this quantity served as the basis for an earlier proposal by Brown et al known as CV-duality \cite{Stanford:2014jda}.\\

Taking inspiration from CV-duality and CA-duality, we would like to propose now that the complexity is dual to the spacetime volume of the WDW patch (more precisely the spacetime volume multiplied by the pressure):
\begin{equation}
    \mathcal{C} \sim \frac{1}{\hbar} P \mathrm{(Spacetime~Volume)}
\end{equation}
In the late time regime, by design we will have:
\begin{equation}
    \dot{\mathcal{C}} \sim \frac{PV}{\hbar}
\end{equation}
We will refer to this proposal as ``complexity=volume 2.0''. 

Next, we ask the question of whether ``complexity=volume 2.0'' satisfies the Lloyd bound. Naively, it might seem that the Lloyd bound favors CA-duality over our proposal, because we have the mass $M$ coming out of CA-duality calculation as opposed to $PV$, and the Lloyd bound refers to the energy of the system.

However, we can form 3 quantities with the dimension of energy out of the standard thermodynamical variables, by multiplying each variable by its conjugate. Thus we have: $M$, $TS$ and $PV$. While $M$ seems to be the ``correct'' energy from the viewpoint of the Lloyd bound, it is $TS$ which should be identified as the complexification rate from the viewpoint of quantum circuits \cite{Stanford:2014jda}. To see this, \cite{Stanford:2014jda} argues that if we think about the CFT as a quantum circuit of $K$ qubits, then the complexity grows linearly in time with slope $K$:
\begin{equation}\label{KdotTime}
    \mathcal{C}{(t_{L},t_{R})} = K|t_{L}+t_{R}|
\end{equation}
To convert between the quantum circuit picture and the field theory picture, we identify $K$ with the entropy $S$ of the CFT and use the temperature $T$ to convert between the CFT time and the quantum circuit time. Thus, we find after the translation:

\begin{equation}
    \mathcal{C}{(t_{L},t_{R})} \sim TS|t_{L}+t_{R}|
\end{equation}

and

\begin{equation}\label{CdotasTS}
    \dot{\mathcal{C}} \sim TS
\end{equation}

On the other hand, one could make similar arguments to make the case that the complexification rate should be $PV$. The complexity should again be proportional to the number of degrees of freedom, which for a discretized CFT is roughly the central charge times the number of lattice sites. Now by the holographic dictionary, we know that the central charge is dual to what we have been calling the pressure. For example, in three bulk dimensions we have the Brown-Henneaux formula \cite{Brown:1986nw}
\begin{equation}
c = \frac{3 L}{2 G} \propto P^{-\frac{1}{2}}
\end{equation}
It furthermore seems reasonable that the volume would roughly encode the number of sites. Thus, one can schematically write down:
\begin{equation}
\mathcal{C} = PV (t_L + t_R)
\end{equation}
and the complexification rate at late time is $PV$.

We end this section by noting that for most black holes all three quantities $M$, $TS$ and $PV$ have the same order of magnitude. To see this, we express these quantities as functions of $r_{+}$ and $L$:
\begin{equation}
    M = \frac{r_{+}}{2} \left( 1 + \frac{r_{+}^{2}}{L^{2}} \right)
\end{equation}
\begin{equation}
    TS = \frac{r_{+}}{4} \left( 1 + \frac{3r_{+}^{2}}{L^{2}} \right)
\end{equation}
\begin{equation}
PV = \frac{r_{+}^{3}}{2L^{2}}
\end{equation}
For $r_{+} >> L$ (i.e. large black holes), we then find
\begin{equation}
    M \approx \frac{r_{+}^{3}}{2L^{2}}
\end{equation}
\begin{equation}
    TS \approx \frac{3r_{+}^{3}}{4L^{2}}
\end{equation}
Interestingly, the two quantities $M$ and $TS$ differ by an $\mathcal{O}{(1)}$ numerical factor, while $M$ and $PV$ become the same quantity! Thus, for high temperatures at least, it does not make much of a difference whether the rate of growth of the complexity is thought of as $M$, as $TS$ or as $PV$. Given that there are ambiguities associated with defining the complexity (such as overall numerical factors), the discrepancies between $M$, $TS$ and $PV$ seem relatively easy to accommodate.

\section{Thermodynamic Volume and Complexity: Conserved Charges}\label{Sec:ComplexityII}
Given the clean connection between the Schwarzschild-AdS WDW patch with the thermodynamic volume and the complexity, it is natural to ask whether we can also establish similar connections for charged and rotating solutions. Unfortunately, within the framework of CA-duality, the situation for charged and rotating black holes is not as clean, and the gravity calculation does not respect the Lloyd bound. In this section, however, we will simply present the computation of the complexity according to ``complexity=volume 2.0'' for a variety of charged black holes, and demonstrate that - like in the uncharged case - the thermodynamic volume and the pressure emerge naturally from the late-time rate of growth. We will relegate the interesting question of consistency with the Lloyd bound to the next section.

\begin{figure*}[h!]
$$
 \includegraphics[width=7cm]{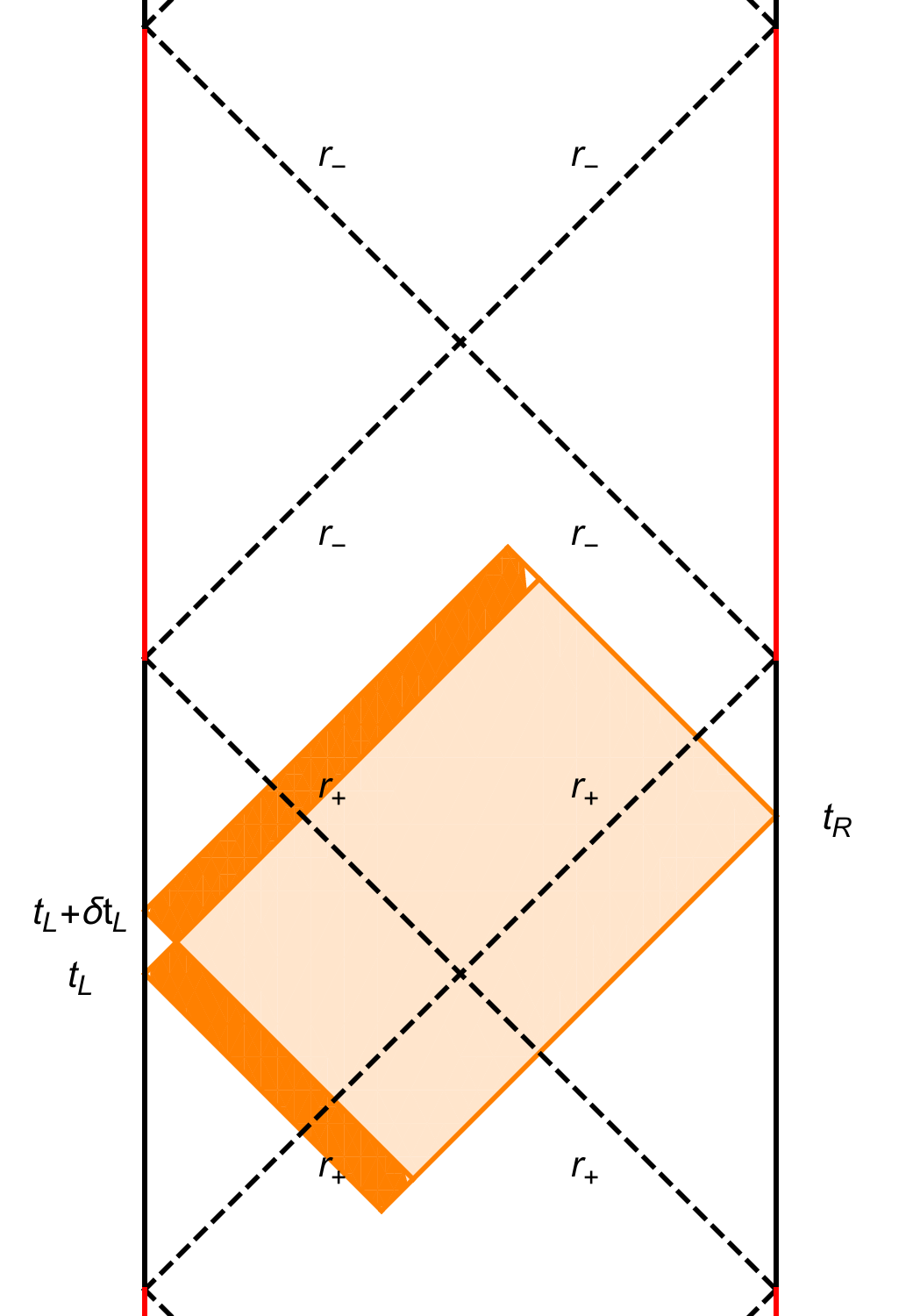}
$$
\caption{The Penrose diagram of a charged and/or rotating black hole and a Wheeler-DeWitt patch (depicted in orange). When $t_{L}$ is shifted to $t_{L} + \delta t_{L}$, the patch loses a sliver and gains another one (depicted in darker orange). The singularity is in red, and the horizons are dashed.}
\label{Fig:WdWpatchAdSRN}
\end{figure*}

On the gravity side, for both charged and rotating black holes, the Penrose diagram is qualitatively the same. In Figure \ref{Fig:WdWpatchAdSRN}, we depict their Penrose diagram together with the WDW patch. Note that the WDW patch is qualitatively different from that of the Schwarzschild-AdS solution: the upper part of the patch no longer runs into a singularity, but approaches the inner horizon at late time.\\

\subsection{Electrically charged black holes}
Let us start with the Reissner-Nordstr\"{o}m black hole in $n+2$ dimensions. The metric together with the gauge field are given by:
\begin{equation}
ds^{2} = -f{(r)}dt^{2} + \frac{dr^{2}}{f{(r)}} + r^{2} d\Omega_{n}^{2}
\end{equation}
\begin{equation}
f{(r)}) = 1 - \frac{\omega^{n-1}}{r^{n-1}} + \frac{q^{2}}{r^{2(n-1)}} + \frac{r^{2}}{L^{2}}
\end{equation}
\begin{equation}
A = \sqrt{\frac{n}{2(n-1)}} \left( \frac{q}{r_{+}^{n-1}} - \frac{q}{r^{n-1}} \right) dt
\end{equation}
As mentioned in the introduction, the thermodynamic volume is well-known and looks like the geometric volume of a ball in flat space:
\begin{equation}\label{VolumeAdSRN}
V_{\pm} = \frac{\mathrm{Vol}{(S_{n})}}{n+1} r_{\pm}^{n+1}
\end{equation}
where the subscript $\pm$ of course refers to either horizon. The spacetime volume of the WDW patch takes the form:
\begin{equation}\label{SpacetimeVolumeAdSRN}
\mathrm{Spacetime~volume} = \frac{\mathrm{Vol}{(S_{n})}}{n} (r_{+}^{n}-r_{-}^{n})(t_{L}+t_{R}) + \dots
\end{equation}
where the ellipsis stand for terms which are time-independent (and therefore drop out from the time derivative of the complexity) or are exponentially suppressed at late time. We recognize the difference between thermodynamic volumes in the equation above. At late time, then, we have as advertised:
\begin{equation}\label{Cdot2horizons}
\lim_{t_{L} \rightarrow \infty} \dot{\mathcal{C}} = \frac{P (V_{+} - V_{-})}{\hbar}
\end{equation}
Note the slight difference compared with the Schwarzschild-AdS case: the late-time complexification rate is now proportional to the \textit{difference} between the two thermodynamic volumes. Let us end this subsection by mentioning the 3-dimensional case of the charged BTZ black hole. Here there is potential for some surprise, since the volume takes the somewhat different form:
\begin{equation}\label{VolumeChargedBTZ}
    V_{\pm} = \pi r_{\pm}^{2} - \frac{\pi}{4} Q^{2} L^{2}
\end{equation}
But in the end, the second term on the right-hand side above drops out of the difference $V_{+}-V_{-}$, and the late-time rate of change of the complexity still takes the form (\ref{Cdot2horizons}).

\subsection{Rotating Black Hole}
\begin{figure*}[h!]
$$
 \includegraphics[width=12cm]{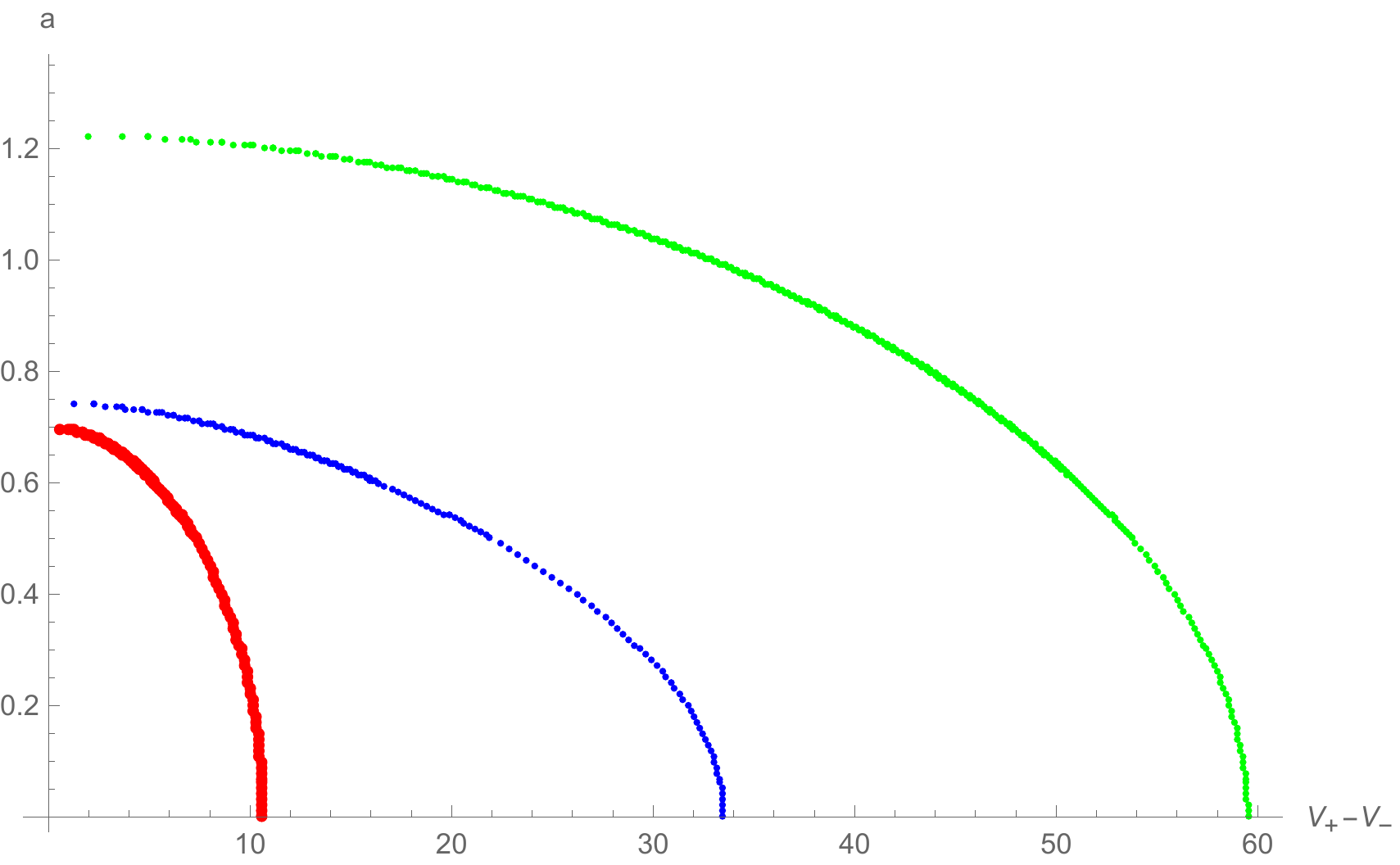}
$$
\caption{Given $M$ and $L$, we vary the angular momentum to mass ratio $a$ and for each value solve numerically for $V=V_+-V_-$. Notice that $a=0$, which reduces to the Schwarzschild case, has the maximal $V$. As we approach extremality, which here occurs as the plots flatten out on the left (In flat space exremality occurs for $a = 1$, but this is modified by the AdS length dependence of the metric), $V$ tends towards zero. In the plot green is for $M=5,L=1$, blue is for $M=2,L=3$, and red is for $M=1,L=2$. }
\label{Fig:J vs V}
\end{figure*}

Next, we move on to discuss rotating holes. Like the Schwarzschild-AdS case, rotating holes are vacuum solutions of the Einstein-Hilbert action, and this again implies that the on-shell Einstein-Hilbert action (ignoring boundary contributions) is proportional to the pressure multiplied by the spacetime volume of the WDW patch:
\begin{equation}
S_{\mathrm{Einstein-Hilbert}} \propto \frac{1}{L^{2}} \int d^{n}x \sqrt{-g} \propto P \mathrm{(Spacetime~volume)}
\end{equation}
Thus, like for the Schwarzschild-AdS case, the distinction between the bulk action (i.e. without the boundary term) and spacetime volume is not very important here. In the simple case of the rotating BTZ black hole, the metric reads:
\begin{equation}
ds^{2} = -f{(r)}dt^{2} + \frac{dr^{2}}{f{(r)}} + r^{2} \left(d\phi - \frac{J}{2r^{2}}dt \right)^{2}
\end{equation}
The thermodynamic volume can be found to be (see for example \cite{Frassino:2015oca} for the outer horizon volume):
\begin{equation}
V_{\pm} = \pi r_{\pm}^{2}
\end{equation}
After some calculation, the late time rate of complexification is again found to be proportional to the difference between the two thermodynamic volumes:
\begin{equation}
\dot{\mathcal{C}} = P(V_{+} - V_{-})
\end{equation}
Next, we move on to discuss the case of rotating black hole in higher dimensions (the Kerr-AdS). This case is substantially richer and more interesting, as the analysis of the thermodynamics is somewhat different depending on whether the spacetime dimension is odd or even (see \cite{Cvetic:2010jb}), and there are two possible notions of volume one can identify. For simplicitly, we will focus on the 4-dimensional case. The solution is given by:
\begin{eqnarray}
ds^{2} &=& -\frac{(1+g^{2}r^{2})\Delta_{\theta}}{1-a^{2}g^{2}} dt^{2} + \frac{(r^{2}+a^{2})\sin^{2}{\theta}}{1-a^{2}g^{2}} d\phi^{2} + \frac{\rho^{2}dr^{2}}{\Delta_{r}} + \frac{\rho^{2}d\theta^{2}}{\Delta_{\theta}} \nonumber \\
&+& \frac{2mr}{\rho^{2}(1-a^{2}g^{2})^{2}} (\Delta_{\theta} dt - a \sin^{2}{\theta} d\phi)^{2}
\end{eqnarray}
\begin{equation}
\Delta_{r} = (r^{2} + a^{2})(1+g^{2}r^{2}) - 2mr
\end{equation}
\begin{equation}
\Delta_{\theta} = 1 - a^{2}g^{2}\cos^{2}{\theta}
\end{equation}
\begin{equation}
\rho^{2} = r^{2} + a^{2}\cos^{2}{\theta}
\end{equation}
Here $a= J/M$ is the ratio of the angular momentum to the mass. The late-time growth of the bulk Einstein-Hilbert action was computed in \cite{Cai:2016xho}:
\begin{equation}\label{LateTimeGrowthKerrAdS}
\frac{d\mathcal{A}}{dt_{L}} = -\frac{1}{2G(L^{2}-a^{2})} \left( r_{+}^{3} + a^{2}r_{+} - r_{-}^{3} - a^{2}r_{-} \right)
\end{equation}
which again is proportional to the spacetime volume by virtue of the solution being a vacuum solution. As for the thermodynamic volume, we have two different notions of volume depending on whether the analysis is done in a non-rotating or rotating frame at infinity. Following \cite{Cvetic:2010jb}, we refer to the volume in the non-rotating frame as the thermodynamic volume and the one in the rotating frame as the geometric volume. The latter admits a geometrical interpretation \footnote{We also note here that the thermodynamic quantities derived in the rotating frame obey the Smarr relation \cite{Cvetic:2010jb} but not the first law. On the other hand, the thermodynamic quantities derived in the non-rotating frame at infinity do obey a first law (in addition to a Smarr relation) and can be derived from the Iyer-Wald formalism.}:
\begin{equation}
V_{+} = \frac{1}{3}r_{+}A_{+}
\end{equation}
where $A$ is the area of the horizon:
\begin{equation}
A_{+} = 4\pi \left( \frac{r_{+}^{2}+a^{2}}{1-a^{2}/L^{2}} \right)
\end{equation}
Putting the two equations above together, we have:
\begin{equation}\label{VoutKerrAdS}
V_{+} = \frac{4}{3}\pi r_{+} \left( \frac{r_{+}^{2}+a^{2}}{1-a^{2}/L^{2}} \right)
\end{equation}
As in the previous cases, we can define a second volume $V_{-}$ associated to the inner horizon by the replacement $r_{+} \rightarrow r_{-}$ in $V_{+}$:
\begin{equation}\label{VinKerrAdS}
V_{-} = \frac{4}{3}\pi r_{-} \left( \frac{r_{-}^{2} + a^{2}}{1 - a^{2}/L^{2}} \right)
\end{equation}
Comparing equations (\ref{LateTimeGrowthKerrAdS}) and (\ref{VoutKerrAdS}), and converting from the bulk action to the spacetime volume, we finally find:
\begin{equation}
\dot{\mathcal{C}} = P(V_{+} - V_{-})
\end{equation}
To help gain intuition, in Figure \ref{Fig:J vs V}, we plot the angular momentum-to-mass ratio $a$ versus $V_{+}-V_{-}$ for fixed $M$ and $L$.

\section{Action or Volume?}\label{ActionvsVolume}

In this paper, we have discussed two different possible holographic identifications of the complexity of the boundary thermofield double state. On could identify this complexity on the one hand with the action of the Wheeler-DeWitt patch, and on the other hand with the spacetime volume of the same. These two quantities behave in a rather similar fashion, and one is naturally led to ask whether any advantage can be identified for one or the other. One advantage is that there are no boundary terms, which in higher curvature theories could have problems near a singularity.\footnote{We thank Adam Brown for pointing this out to us.}. In this section, we seek to answer this question as regards the Lloyd bound \cite{Lloyd}.

\subsection{The Lloyd Bound with conserved charge}

In this subsection, let us derive the Lloyd bound in the presence of a conserved charge. As argued in \cite{Brown:2015lvg}, the existence of conserved charges puts constraints on the system and implies that the rate of growth of the complexity at late time is slower than in the case without charges. Let us start by generalizing the thermofield double state to include a chemical potential $\mu$:
\begin{equation}
  | TFD_{\mu} \rangle = \frac{1}{\sqrt{Z}} \sum_{n} e^{-\beta (E_{n}+\mu Q_{n})/2} |E_{n} Q_{n} \rangle_{L} | E_{n} -Q_{n} \rangle_{R}
\end{equation}
This state time-evolves by the Hamiltonian $H_{L} + \mu Q_{L}$ on the left, and $H_{R} - \mu Q_{R}$ on the right: \footnote{Note the difference in the sign of $\mu$ between $H_{L}$ and $H_{R}$. This is because the electrostatic potential is positive on one side and negative on the other.}
\begin{equation}
| \psi{(t_{L},t_{R})} \rangle = e^{-i(H_{L} + \mu Q_{L}) t_{L}} e^{-i(H_{R} - \mu Q_{R}) t_{R}} |TFD_{\mu} \rangle
\end{equation}
Based on this, one would guess that the appropriate generalization of the Lloyd bound is:
\begin{equation}
\dot{\mathcal{C}} \leq  \frac{2}{\pi \hbar}\left(M - \mu Q \right)
\end{equation}
This however violates our intuition that as zero temperature system, the complexification rate of an extremal black hole should be zero, and that the bound should reflect this. It thus seems appropriate to modify the above to

\begin{equation}
\dot{\mathcal{C}} \leq  \frac{2}{\pi \hbar}\left[ \left(M - \mu Q \right) - \left( M - \mu Q \right)_\text{gs} \right]
\end{equation}

where $\left( M - \mu Q \right)_\text{gs}$ is nothing but $M - \mu Q$ evaluated in the appropriate ground state, which will be either empty AdS or an extremal black hole depending on the case under consideration. If we think of our system as being in the grand canonical ensemble, it is most natural to take the ground state to correspond to the geometry whose chemical potential is the same as the black hole under consideration. As it happens, this is nothing but pure AdS for black holes with $\mu \leq 1$, but for $\mu > 1$, it corresponds to some extremal black hole (In units where $G=1$). 

\subsection{Bound Violation: Near the Ground State}

Now we will check to see whether the Lloyd bound is obeyed by the two proposals at hand. For simplicity, we restrict our attention to 4-dimension and work in units where $G=1$. First, we consider the case where $\mu>1$. Expanding the outer horizon radius near extremality, we find that 

\begin{equation}
r_+ \approx r_e \left( 1 + \frac{\sqrt{3}}{\mu ^2 \sqrt{\mu ^2-1} L} \delta M + \mathcal{O}(\delta M^2) \right)
\end{equation}

Where $\delta M := M - M_e$, $M$ is the total mass of the black hole, and $r_e$ and $M_e$ are the radius and mass respectively of an extremal black hole with the same chemical potential as the one we are considering. We likewise may expand the inner horizon as

\begin{equation}
r_- \approx r_e \left( 1-\frac{\sqrt{3}}{\mu ^2 \sqrt{\mu ^2-1} \left(2 \mu ^2-1\right) L} \delta M + \mathcal{O}(\delta M^2)\right)
\end{equation} 

From these, we can expand $\dot{C}$ under both proposals as

\begin{equation}
\dot{C}_V = P \left(r_+^3 - r_-^3 \right) \approx r_e^3 \left[ \frac{9 \sqrt{3} \sqrt{\mu ^2-1}}{4 \pi  \mu ^2 \left(2 \mu ^2-1\right) L^3} \delta M + \mathcal{O}(\delta M^2) \right]
\end{equation}

\begin{equation}
\dot{C}_A = \frac{Q^2}{r_-} - \mu Q \approx \frac{2 \left(\mu ^2-1\right)}{2 \mu ^2-1} \delta M + \mathcal{O}(\delta M^2)
\end{equation}

On the other hand, the bound is given by

\begin{equation}
(M - \mu Q) - (M - \mu Q)_e \approx \frac{\sqrt{3} \sqrt{\mu ^2-1}}{2 \mu ^4 L} \delta M^2 + \mathcal{O}(\delta M^3)
\end{equation}

We thus see that both proposals must violate the Lloyd bound near extremality. This is in agreement with \cite{Brown:2015lvg}.

Next we consider $\mu \leq 1$, in which case we expand around empty AdS. Here the bound becomes to lowest order

\begin{equation}
2 \left( M - \mu Q \right) \approx \left(1-\frac{2 \mu ^2}{\mu ^2+1}\right) M = 2 \left( \frac{1-\mu^2}{1+\mu^2} \right) M
\end{equation}

But $\dot{C}$ becomes under each proposal

\begin{equation}
\dot{C}_V \approx \frac{3 \left(1 - \mu ^6\right)}{\pi  \left(\mu ^2+1\right)^3 L^2} M^3 + \mathcal{O}(M^5)
\end{equation}

\begin{equation}
\dot{C}_A \approx 2 \left( \frac{1-\mu^2}{1+\mu^2} \right) M + \mathcal{O}(M^3)
\end{equation}

We see immediately that the bound is satisfied in CV-duality sufficiently near extremality, but the bound is far from saturated.The situation with CA-duality is a bit more complex: $\dot{C}_A$ exactly saturates the bound to lowest order in $M$, and so the lower order behavior becomes important. Expanding the bound violation (i.e. the difference between $\dot{C}_A$ and the bound) directly we find to lowest order

\begin{equation}
\dot{C}_A - 2 \left( M - \mu Q \right) \approx \frac{8 \mu ^2}{\left(\mu ^2+1\right)^2 L^2} M^3 + ...
\end{equation}

As this term is positive definite, we see that the bound is violated as we approach empty AdS. This would seem to put CV-duality in a slightly better position than CA-duality, though the expectation that the bound should be saturated, or nearly so is not met in this case.

\subsection{Bound Violation: Exact Results}

In the $\mu\leq 1$ case, one can, in fact, do better. We can find the bound violations as an exact function of the inner and outer horizon. Note of course that these are only valid over the region in $r_+,r_-$ space where $\mu\leq 1$. The exact expressions in 4 dimensions are

\begin{equation}
\dot{C}_V - 2(M - \mu Q) = \frac{2 L^2 (r_--r_+)+r_-^3+2 r_-^2 r_++2 r_- r_+^2-r_+^3}{2 L^2}
\end{equation}

$$ = -\frac{(r_+ - r_-)^3 + 2 L^2 (r_+ - r_-) + r_+ r_- (r_+ - 5 r_-)}{2L^2}$$

and

\begin{equation}
\dot{C}_A - 2(M - \mu Q) = \frac{r_+ r_- (r_+ + r_-)}{L^2}
\end{equation}

The second expression is clearly positive definite, and so the under CA-duality the bound is always violated for $\mu \leq 1$. Now the exact expression for the chemical potential is 

\begin{equation}
\mu = \sqrt{\frac{r_- (L^2 + r_-^2 + r_- r_+ + r_+^2) }{r_+ L^2}}
\end{equation}

From this we may conclude that 

\begin{equation}\label{mulessthan1}
\mu^2 \leq 1 \Rightarrow r_- (r_+^2 + r_+ r_- + r_-^2) \leq L^2 (r_+ - r_-)  
\end{equation}

and so 

\begin{equation}
\dot{C}_V - 2(M - \mu Q) \leq \frac{-(r_+ - r_-)^3 - 2 r_- (r_+^2 + r_+ r_- + r_-^2) - r_+ r_- (r_+ - 5 r_-)}{2L^2}
\end{equation}

$$ = - \frac{r_+^3 + r_-^3}{2L^2} \leq 0$$

and so we may conclude that $C_V$ respects the bound whenever $\mu \leq 1$.

Generalizing further to arbitrary dimension $d>3$, we find that 

\begin{equation}
\dot{C}_A - 2(M - \mu Q) = \frac{(d-2) \Omega_d r_+^d r_-^d (r_+^2 - r_-^2)}{8 \pi L^2 (r_+^d r_-^3 - r_+^3 r_-^d)}
\end{equation}
Which is a positive definite quantity. This in fact recovers a result already derived by \cite{Cai:2016xho}. Being, for now, a bit less ambitions with the $CV$ quantity, we find in 5 dimensions
\begin{equation}
\dot{C}_V - 2(M - \mu Q) = - \frac{3 \pi  \left(2 L^2 \left(r_+^2 - r_-^2\right) + r_+^4 - 2 r_-^2 r_+^2 - r_-^4 \right)}{8 L^2}
\end{equation}
and 
\begin{equation}
\mu^2 = \frac{3 \pi  r_-^2 \left(L^2+r_-^2+r_+^2\right)}{4 L^2 r_+^2 } \leq 1 \Rightarrow r_+^2 \geq \frac{3 \pi  r_-^2 \left(L^2+r_-^2+r_+^2\right)}{4 L^2 }
\end{equation}
from which we get
\begin{equation}
\dot{C}_V - 2(M - \mu Q) \leq - \frac{3 \pi  \left(2 L^2 \left( \frac{3 \pi  r_-^2 \left(L^2+r_-^2+r_+^2\right)}{4 L^2}- r_-^2 \right) + r_+^4 - 2 r_-^2 r_+^2 - r_-^4 \right)}{8 L^2}
\end{equation}
$$ = -\frac{3 \pi  \left((3 \pi -4) L^2 r_-^2+(3 \pi -2) r_-^4+(3 \pi -4) r_-^2 r_+^2+2 r_+^4\right)}{16 L^2} \leq 0$$
And so CV duality in 5 dimensions respects the bound whenever $\mu^2 < 1$. We conjecture without proof that CV-duality repects the Lloyd bound whenever $\mu^2\leq1$ for all AdS-RN spacetimes.

\subsection{Altering the Bound by a Pre-Factor}

We have considered the Lloyd bound in it's usual form, 

\begin{equation}
\dot{C} \leq \frac{2 E}{\pi \hbar}.
\end{equation}

It would seem, however, due to the arguments leading to this bound, that the bound should only be trusted up to an overall factor. It would be interesting, therefore, to see how robust the above discussion is under the insertion of some pre-factor. For example, under which proposals and sets of circumstances does
\begin{equation}
\dot{C} \leq \frac{\alpha E}{\pi \hbar}.
\end{equation}
hold for various values of $\alpha$. For example, for $\alpha=1$ for the AdS-RN case we find:
\begin{equation}
    \dot{C}_{V} - (M - \mu Q) = \frac{r_{+}r_{-}(r_{+} + r_{-}) - L^{2}(r_{+}-r_{-})}{2L^{2}}
\end{equation}
and, using the inequality (\ref{mulessthan1}) again, we find:
\begin{equation}
    \dot{C}_{V} - (M - \mu Q) \leq -\frac{r_{-}^{3}}{2L^{2}} < 0
\end{equation}
Hence the value $\alpha=1$ is also consistent with the bound. We leave further explorations of other values of $\alpha$ to future research.

\section{Conclusion}\label{Sec:Conclusion}
Let us summarize the main findings in this paper and sketch out some future directions. In the first part of the paper, we analyzed the notion of thermodynamic volume from the viewpoint of the Iyer-Wald formalism. Using a slight generalization this formalism, we present a systematic way to derive the volume and illustrate it in two cases: the charged BTZ black hole and the R-charged black hole. In the latter case, our method explains several interesting and intriguing features of the thermodynamic volume, and we believe that it will prove useful to compute the volume of many more complicated black hole solutions in the future. Of particular interest are Lifschitz black holes \cite{Liu:2014dva, Brenna:2015pqa}. Even though the computation of the volume for the R-charged black hole was a bit involved, it is still relatively simple since we saw that perturbing the coupling $g$ leaves all the matter fields unchanged. In comparison, we do not have this luxury in the case of Lifschitz solutions: a generic feature of these spacetimes is the fact that the profiles of the matter fields depend explicitly on the cosmological constant (this property is somehow related to the fact that these spacetimes are not asymptotically AdS), so varying the cosmological constant will affect the matter fields.

In the second part of the paper, we related the thermodynamic volume to the holographic proposals for the complexity. In particular, we showed that the thermodynamic volume (together with its conjugate the pressure) is intimately related to the WDW patch of an eternal AdS black hole, and this holds for a large class of AdS black holes. This intimate relationship can be stated cleanly in two different ways: On one hand, the rate of change of the WDW spacetime volume in the late time limit is precisely the thermodynamic volume (if there is only one horizon) or the difference of thermodynamic volumes (if there are two horizons). On the other hand, the bulk action evaluated on the WDW patch (ignoring boundary contributions) is the sum of ``work terms'' involving pressure-volume and charge-potential. The several different ways to arrive at the thermodynamic volume presented in this paper may be a little confusing to the reader, so let us state again the relationship between them: The thermodynamic volume may be defined in the usual thermodynamic fashion as the partial derivative of the ADM mass with respect to the pressure. The volume computed by the Iyer-Wald formalism is by construction the same quantity. We conjecture that this should further correspond to the late-time value of the time derivative of the spacetime volume of the Wheeler-DeWitt patch, and have checked several examples, but have no proof that this holds generally. 

How to take this story further? As mentioned in the introduction, a tensor network picture of the black hole interior was introduced by Hartman and Maldacena in \cite{Hartman:2013qma}. Tensor network is a topic of much recent interest for holographers \cite{Czech:2015qta,Czech:2014tva,Czech:2015B,Swingle:2009bg} with an eye on the emergence of spacetime. Thus one can ask the question: can the pressure-volume variables be understood in the language of tensor networks or quantum circuits? Also, according to black hole complementarity \cite{Susskind:1993if,'tHooft:1984re}, the black hole interior is an example of emergent space {\it par excellence}. Thus, one can hope that the pressure and volume variables can prove helpful to our understanding of quantum gravity in the future. 

\section{Acknowledgement}
We would like to thank Adam Brown, Bartlomiej Czech, and Leonard Susskind for the useful comments which they provided on an early draft of this paper. We would further like to thank Kimberly Carmona for assisting in proof-reading the manuscript. This material is based upon work supported by the National Science Foundation under Grant Number PHY-1620610.

\end{document}